\documentclass[journal]{IEEEtran}
\usepackage{color}
\usepackage[table,xcdraw]{xcolor}
\pdfoutput=1

\usepackage{cite}
\ifCLASSINFOpdf
   \usepackage[pdftex]{graphicx}
\else
   \usepackage[dvips]{graphicx}
\fi
\usepackage{amsmath}
\DeclareMathOperator{\sign}{sgn}
\ifCLASSOPTIONcompsoc
  \usepackage[caption=false,font=normalsize,labelfont=sf,textfont=sf]{subfig}
\else
  \usepackage[caption=false,font=footnotesize]{subfig}
\fi

\usepackage{stfloats}
\begin{document}
\title{Three-Dimensional GPU-Accelerated Active Contours for Automated Localization of Cells in Large Images}

\author{Mahsa~Lotfollahi\textsuperscript{1}, Sebastian~Berisha\textsuperscript{1}, Leila~Saadatifard\textsuperscript{1}, Laura~Montier\textsuperscript{2},  Jokubas~Ziburkus\textsuperscript{2}, and David~Mayerich\textsuperscript{1}*% <-this % stops a space
\thanks{M. Lotfollahi, S. Berisha and D. Mayerich are with the Department of Electrical and Computer engineering, University of Houston, Houston, TX.
}% <-this % stops a space
\thanks{L. Montier and J. Ziburkus are with the department of Biology and Biochemistry, University of Houston, TX}}% <-this % stops a space
\maketitle

\begin{abstract}
Cell segmentation in microscopy is a challenging problem, since cells are often asymmetric and densely packed. This becomes particularly challenging for extremely large images, since manual intervention and processing time can make segmentation intractable. In this paper, we present an efficient and highly parallel formulation for symmetric three-dimensional (3D) contour evolution that extends previous work on fast two-dimensional active contours. We provide a formulation for optimization on 3D images, as well as a strategy for accelerating computation on consumer graphics hardware. The proposed software takes advantage of Monte-Carlo sampling schemes in order to speed up convergence and reduce thread divergence. Experimental results show that this method provides superior performance for large 2D and 3D cell segmentation tasks when compared to existing methods on large 3D brain images.
\end{abstract}

\begin{IEEEkeywords}
Active contour, snake, cell segmentation.
\end{IEEEkeywords}

% \fi
\IEEEpeerreviewmaketitle

\section{Introduction}
\IEEEPARstart{Q}{uantifying} the size and distribution of cell nuclei in optical images is critical to understanding the underlying tissue structure \cite{su_interactive_2016} and  organization \cite{irshad2014methods , lin_hybrid_2003}. Segmentation is crucial to this analysis, by providing quantitative data that pathologists can use to characterize diseases and evaluate their progression\cite{xing_automatic_2016}. Since manual analysis of microscopy images is time consuming and labor intensive, automated cell segmentation is essential for detecting and locating cells in massive images.
Microscopy images exhibit a large degree of variability and complexity, due to large numbers of overlapping cells and variations in cell types and stages of cell division, imaging systems, and staining protocols. In order to deal with this complexity, a large number of segmentation algorithms have been proposed \cite{wahlby_combining_2004,meijering_cell_2012}. Most current algorithms use basic techniques combined with complicated pipelines to overcome those challenges. These methods include thresholding \cite{xiaobo_zhou_novel_2009, george_automated_2013,xu_efficient_2014}, feature extraction \cite{grigorescu_comparison_2002 , ongun_feature_2001}, classification \cite{yin_cell_2010}, c-means\cite{george_automated_2013} and k-means\cite{jorgensen_using_2017} clustering, region growing \cite{jun_tang_color_2010, koyuncu_smart_2012 , maycock_structure_1976}, and deformable models\cite{sadeghian_framework_2009 , ko_automatic_2011, lotfollahi_segmentation_2017}.

Recently, learning based approaches using artificial neural networks (ANN) and convolutional neural networks (CNN) have gained increased attention. These methods rely on example data to train a machine learning algorithm to identify boundary pixels \cite{ronneberger2015u} or directly perform binary segmentation \cite{chen2016dcan}. In general, most pipelines include prepossessing, finding cell bounding boxes, extracting either spatial or frequency-based features \cite{wang2017accurate,irshad2014methods,akram2016cell,shan2012completely} or using several convolution layers followed by max-pooling \cite{khoshdeli2017detection,sailem2017discovery}, and finally classifying the image. In these approaches, the training phase is time consuming and requires massive amounts of labeled data \cite{cao2017breast}.

Most current algorithms focus on two-dimensional data, such as histology slides, and utilize a variety of techniques to deal with specific tissue types, stains, and labels. For example, deep CNNs have been used for segmentation of overlapping clumps in Pap smear images \cite{song2017accurate}, and support vector machines (SVMs) have been employed to segment epithelial cells \cite{santamaria2015cell} and skeletal muscle \cite{janssens2013charisma}. Finally, active contours have been shown to be effective for cell nuclei segmentation \cite{nielsen2015optimizing}.

The major limitation of histology slices is that they are limited to 2D sampling. Although histological assesments convey some of the structure and morphology of the tissue, they do not provide proper insights into the 3D layout of cells. In addition, 3D images provide much better separability when cells are overlapping or hidden in the corresponding 2D images. 

To date, several software solutions are available for specialized cell segmentation on 3D images. FARSIGHT \cite{al-kofahi_improved_2010} uses graph cuts and multi-scale Laplacian of Gaussian filters to detect cell seed points. Region growing is then used based on local-maximum clustering. MINIS \cite{lou_rapid_2014} performs blob detection by smoothing the image with Gaussian kernels at different scales and computing eigenvalues of the Hessian matrix at each pixel from these smoothed images. It then thresholds the respective eigenvalues to obtain a mask of nuclei and a connected component analysis assigns a unique ID to each nucleus. The 3D object counter plugin for ImageJ \cite{bolte_guided_2006} is a simple 3D cell counter which uses a user-specified intensity threshold to separate foreground and background, resulting in an over-segmented image. Since fundamental thresholding is not robust, adaptive and iterative thresholding on smoothed 3D images can also be used \cite{hodneland_cellsegm_2013,cheng2017adaptive}.

One method for addressing large-scale segmentation relies on simple active contours, such as snakuscules \cite{thevenaz_snakuscules_2008,pediredla_unified_2012}, which are fast to evaluate and rely on very few user-specified input parameters. However, a three-dimensional application of this algorithm has not been derived. In addition, the sampling required to evolve a primitive active contour is computationally intractable for images containing thousands of cells.
 
\section {Approach}
Most deformable models transform an image segmentation task to an optimization problem. An energy function is defined based on the image content and desired behavior of a curve. Snakuscules, introduced in \cite{thevenaz_snakuscules_2008}, are region-based snakes optimized for the segmentation of approximately circular features. In this section, we will describe the previously published snakuscule algorithm as well as generalize the mathematics to three-dimensional images.

\subsection{Snakuscules}
Snakuscules are active contours optimized for fast convergence around circular image features. Their fast evaluation time allows the initialization of many contours that cover an entire image, allowing segmentation of blob-like features without manual initialization.

\begin{figure}
    \centering
    \includegraphics [width=0.9\linewidth]{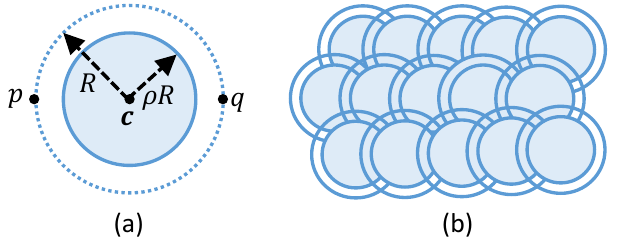}
    \caption{(a) A snakuscule is defined by by two points $\mathbf{p}$ and $\mathbf{q}$. (b) Initial configuration of multitude snakuscules congregated together at a distance $\sqrt{1.5}R$}.
    \label{fig:snake2D-swarm}
\end{figure}

A Snakuscule is defined by a pair of concentric disks parameterized by two points $\mathbf{p}$ and $\mathbf{q}$ (Figure \ref{fig:snake2D-swarm}a). The optimization attempts to minimize an energy function measuring the contrast between the inner disk and annulus in order to completely surround a bright round object with a circular curve. The energy function is defined to balance the weighted inner area against the weighted outer area of the curve:
\begin{align}
\mathbf{E(\mathbf{p} , \mathbf{q})} &= \iint\limits_{\rho R<||\mathbf{x} - \mathbf{c}||<R }I(\mathbf{x})d\mathbf{x} - \iint\limits_{||\mathbf{x}- \mathbf{c}||<\rho R }I(\mathbf{x})d\mathbf{x}\\
R &= \frac{1}{2}|| \mathbf{p} - \mathbf{q}||\nonumber\\
\mathbf{c} &=\frac{1}{2}\left(\mathbf{p}+\mathbf{q}\right)\nonumber
\end{align}
where $I(\mathbf{x})$ is a two-dimensional image, $\mathbf{x}=[x_1, x_2]^T$ is an image coordinate, $R$ is the radius of the snake, and $\mathbf{c}$ is its center. The value $\rho=\frac{1}{\sqrt{2}}$, derived previously for the two-dimensional case \cite{thevenaz_snakuscules_2008}, enforces the equal area for both inner disk and outer annulus.

\begin{figure}[t]
\centering
\includegraphics[width=0.9\linewidth]{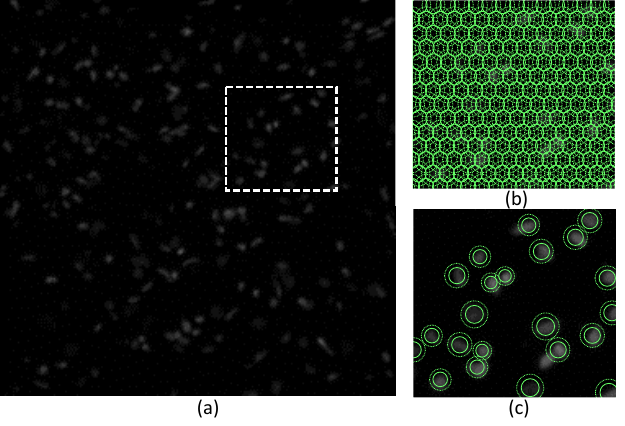}
\caption{ (a) A DAPI stained brain tissue slice. (b) The initial configuration and (c) final configuration of snakuscules on a zoomed region.}
\label{fig:2D conf}
\end{figure}

One snakuscule can find and segment one light blob in the image. To catch all interesting features, many initial contours are specified to cover the image (Figure \ref{fig:snake2D-swarm}b). The contours are then evolved, and trivial contours that do not converge to image features are eliminated (Figure \ref{fig:2D conf}). 

\subsection{3D Snakuscules}
\begin{figure}[t] 
\centering
\includegraphics[width=0.9\linewidth]{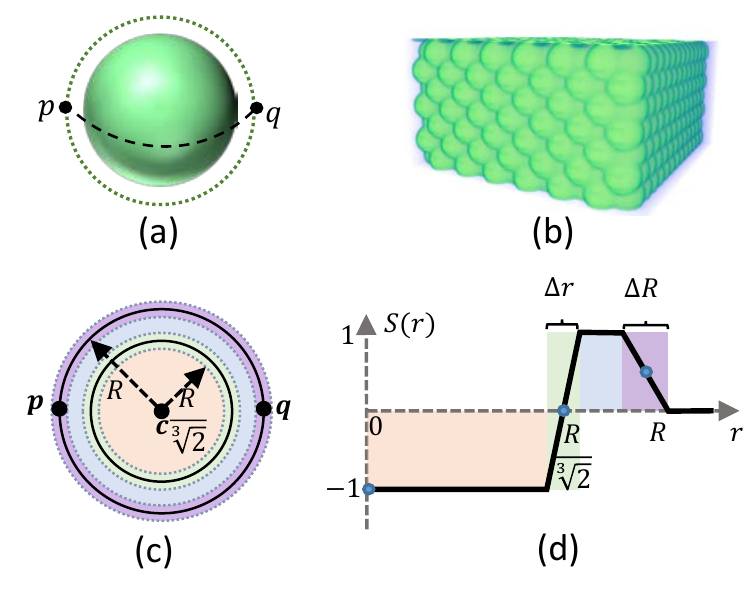}
 \caption{(a) A 3D snakuscule defined by two points $\mathbf{p}$ and $\mathbf{q}$; to simplify computations 3D snakusules identifier points are considered in the same $y$ and $z$ levels. (b) The initial configuration consists of every two neighboring 3D contours located at a distance $\sqrt{1.5}R$ apart. (c) The middle section of the 3D snakuscule with four distinct regions are shown in different colors. (d) The weight function assigns a weight to any portion of the 3D snakuscule shown in (c).}
 \label{fig:all_in_one}
 \end{figure}

We first propose a mathematical framework for evolving snakuscules in 3D by moving and expanding/contracting an initial 3D contour to fit cell nuclei. This generalization makes it viable to extend similar contours to higher-dimensional or hyperspectral data (ex. hypersnakuscules). The 3D snakuscule is based on a pair of concentric spheres that are parameterized by two points $\mathbf{p}=[p_x, p_y, p_z]^T$ an $\mathbf{q}=[q_x, q_y, q_z]^T$ (Figure \ref{fig:all_in_one}a). The optimization process minimizes a local energy function, which favors high contrast between weighted inner and outer volumes.
Contours move and evolve within the spatial domain of an image to minimize the contrast energy function (Equation \ref{eq:general energy}).
\begin{equation}
\mathbf{{E}(\mathbf{p} , \mathbf{q})} = \iiint\limits_{\rho R<||\mathbf{x}- \mathbf{c}||<R } I(\mathbf{x}) d\mathbf{x} - \iiint\limits_{||\mathbf{x}- \mathbf{c}||<\rho R }I(\mathbf{x})d\mathbf{x}
\label{eq:general energy}
\end{equation}

\begin{figure}
    \centering
    \includegraphics[width=0.9\linewidth]{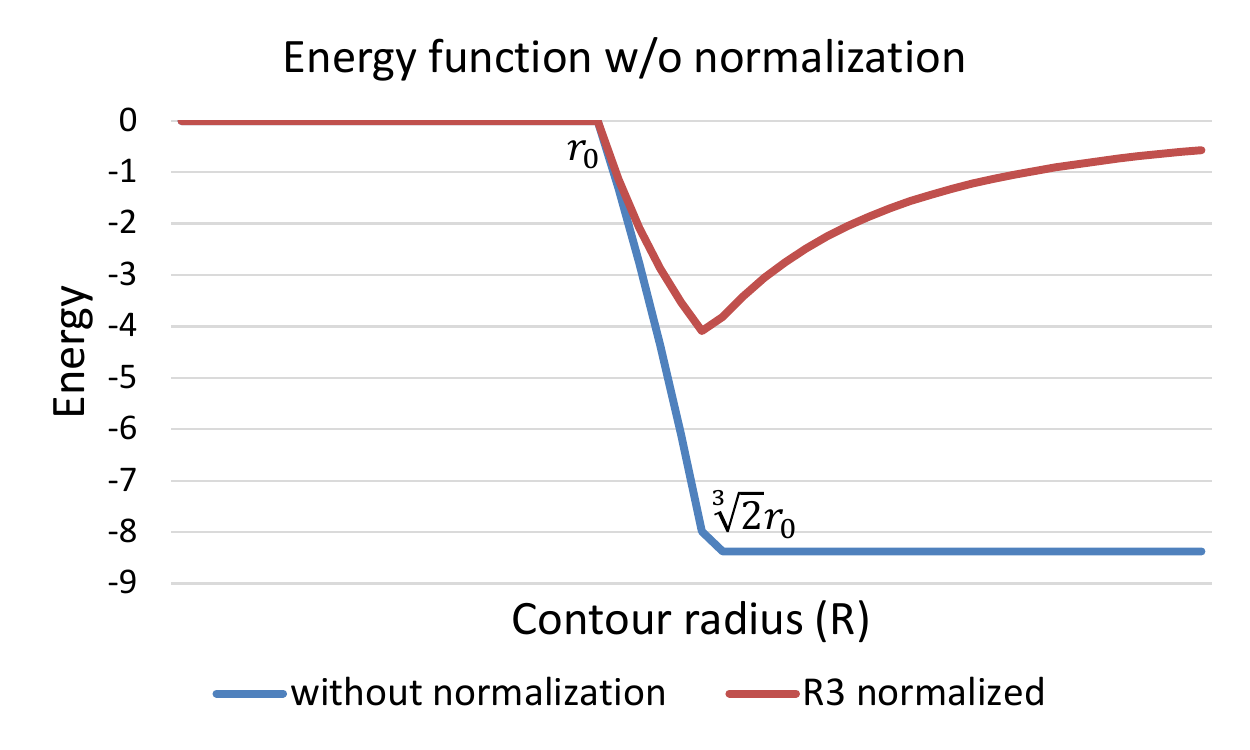}
    \caption{Energy changes of the 3D snake with and without normalization term w.r.t its radius. Energy function with normalization term has a single local minimum when the snake fit the blob.}
    \label{fig:norm-energy}
\end{figure}

This optimization leads the contour toward a bright spherical object on a dark background. To ensure the snake does not move in uniform regions with constant intensity where $\forall \mathbf{x} \in \mathbf{R}^3: \mathbf{I(x)}=\mathbf{I_0}$, the energy is defined using two sub-terms that cancel each other out; therefore, $\rho= \frac{1}{\sqrt[3]{2}}$.
% \begin{equation}
% \mathbf{\hat{E}(\mathbf{p} , \mathbf{q})} = \iiint\limits_{\rho R<||\mathbf{x}- \mathbf{c}||<R } I(\mathbf{x}) d\mathbf{x} - \iiint\limits_{||\mathbf{x}- \mathbf{c}||<\rho R }I(\mathbf{x})d\mathbf{x}
% \label{eq:general energy}
% \end{equation}
% where
% $$\rho= \frac{1}{\sqrt[3]{2}}$$
This prevents the contour from sliding when the surrounding gradient is zero \cite{thevenaz_snakuscules_2008}. We illustrate energy minimization for a generic model of a light blob $I(r,\theta) = 1 + \sign(r_0 - r)$ , which is a sphere of radius $r_0$ in a black background. When the contour is concentric within the blob, the resulting energy is given by:
\begin{equation}
    \mathbf{\hat{E}} = \begin{cases}
    0, & R<r_0\\
    -\frac{8}{3} \pi(R^3 - r_0^3), & \frac{R}{\sqrt[3]{2}}<r_0<R\\
    -\frac{8}{3} \pi r_0^3, & R\geq\sqrt[3]{2}r_0
    \end{cases}
\end{equation}
The energy achieves an optimal value for any contour equal or larger than the blob, so there is no unique optimal contour. To ensure that the volume occupied by the contour is also minimized, a normalization term $\alpha$ is used to reformulate the energy function:
\begin{equation}
\mathbf{\bar{E}}(R,\alpha) = \frac{\mathbf{\hat{E}}}{R^\alpha}
\end{equation}
where $\alpha>0$ to apply a penalty when the contour becomes larger than the blob. To balance the expansion and contraction speed when the contour is approaching the blob size, we force the energy gradient to be symmetric as the optimal value is approached:
$$
\lim_{R\to\sqrt[3]{2} - \epsilon}\frac{\delta \mathbf{\bar{E}}(R,\alpha)}{\delta R} = \lim_{R\to\sqrt[3]{2} + \epsilon}-\frac{\delta \mathbf{\bar{E}}(R,\alpha)}{\delta R}
$$
which results in a normalization value $\alpha = 3$. Figure \ref{fig:norm-energy} depicts the energy function with respect to the contour radius with and without normalization. Note that this normalization term can be optimized as desired for objects that are not binary indicator functions (ex. Gaussian kernels).

%\begin{equation}\label{eq: energy-integ}
%\mathbf{E}=\frac{1}{8R^3}\left[\iiint_{\rho R<||\mathbf{x}- \mathbf{c}||<R }I(\mathbf{x})d\mathbf{x} -\iiint_{||\mathbf{x}- \mathbf{c}||<\rho R }I(\mathbf{x})d\mathbf{x}\right]
%\end{equation}
A discrete formulation of the energy function is generated by substituting summation for integration in the  pixel domain. The final discrete energy function is given by: 
\begin{equation}\label{eq:energy}
\mathbf{E}(\mathbf{p},\mathbf{q}) = \frac{1}{||\mathbf{p}-\mathbf{q}||^3}\sum_{\mathbf{k} \in\mathbf {K}} S(r)I(\mathbf{k})
\end{equation}
where  $\mathbf{K}$ is the set of all pixels within $R+\frac{1}{2}\Delta R$ of the 3D snake center, $r = |\mathbf{k} - \mathbf{c}|$, and $S(r)$ is a differentiable weight function (Figure \ref{fig:all_in_one}d), so that $\int_{0}^{\infty} S(r)r^2dr=0$. The 3D snake is composed of four different regions (Figure \ref{fig:all_in_one}c); two dynamic and two fixed regions. During evolution, the entire footprint becomes smaller or larger while $\Delta R$ and $\Delta r$ remain unchanged:
$$\Delta{r}=\Delta R/\sqrt[3]{2}$$
To simplify calculations, the two identifier points $\mathbf{p}$ and $\mathbf{q}$ are considered to be in the same line along both the $y$ and $z$ directions ($p_y=q_y$ and $p_z=q_z$). The energy function can be rewritten as:
\begin{equation}\label{eq:simple_energy}
\mathbf{E}(\mathbf{p},\mathbf{q}) = \frac{1}{\left (q_x-p_x\right )^3}\sum_{\mathbf{k} \in\mathbf {K}} S(r)I(\mathbf{k})
\end{equation}

\subsection{3D contour evolution}
The 3D snakuscule evolves by movements of $\mathbf{p}$ and $\mathbf{q}$ in the opposite direction of $\nabla \mathbf{E}$ to minimize the energy function using gradient descent. Therefore, partial derivatives of the energy function $\mathbf{E}$ with respect to the identifier points $\mathbf{p}$ and $\mathbf{q}$ are required:
\begin{align}\label{dEdp.x}
\frac{\partial\mathbf{E}}{\partial{p_x}}  &= \gamma\left[\frac{3}{q_x - p_x}\sum\nolimits S(r)I(\mathbf{k}) +\sum\nolimits \frac{\partial{ S}}{\partial{p_x}}I(\mathbf{k})\right]\\
\label{dEdq.x}
\frac{\partial\mathbf{E}}{\partial{q_x}}  &= \gamma\left[\frac{-3}{q_x - p_x}\sum\nolimits S(r)I(\mathbf{k}) + \sum\nolimits \frac{\partial{ S}}{\partial{q_x}}I(\mathbf{k})\right]\\
\label{dEdp.y}
\frac{\partial\mathbf{E}}{\partial{q_y}}&=\frac{\partial\mathbf{E}}{\partial{p_y}}=  \gamma\sum\nolimits\frac{\partial{ S}}{\partial{p_y}}I(\mathbf{k})\\
\label{dEdp.z}
\frac{\partial\mathbf{E}}{\partial{q_z}}&=\frac{\partial\mathbf{E}}{\partial{p_z}}=  \gamma\sum\nolimits\frac{\partial{ S}}{\partial{p_z}}I(\mathbf{k})
\end{align}
where
\begin{equation}
\gamma = \frac{1}{(q_x -p_x)^3}
\end{equation}

% The hypersnakuscule is composed of four different regions (Figure \ref{fig:all_in_one}(e)); two dynamic and two fixed regions. During evolution, the whole footprint becomes smaller or larger while $\Delta R$ and $\Delta r$ remain unchanged, then $\Delta{r}=\Delta R/\sqrt[3]{2}$.

We minimize the energy (Equation \ref{eq:simple_energy}) using gradient descent to update the position of the identifier points.
Each point $\mathbf{k} \in \mathbf{K}$ applies a force to $\mathbf{p}$ and $\mathbf{q}$ that dictate its motion over time:
\begin{align}\label{sum_dEdp}
\frac{d \mathbf{p}}{d t} &= -\sum_{\mathbf{k}\in\mathbf{K}} \frac{\partial \mathbf{E}(\mathbf{k})}{\partial \mathbf{p}}\\
\label{sum_dEdq}
\frac{d \mathbf{q}}{d t} &= -\sum_{\mathbf{k}\in\mathbf{K}} \frac{\partial \mathbf{E}(\mathbf{k})}{\partial \mathbf{q}}\\
\mathbf{p}_{n+1} &= \mathbf{p}_n + \epsilon \frac{d \mathbf{p}}{d t}\\
\mathbf{q}_{n+1} &= \mathbf{q}_n + \epsilon \frac{d \mathbf{q}}{d t}
\end{align}
where $\epsilon=\frac{\epsilon_0}{\sqrt{n}}$ is learning rate, $\epsilon_0$ is constant and n is the iteration number.

% \begin{figure*}[t] 
% \centering
% \includegraphics[width=0.75\linewidth]{figures/swarm-initials.pdf}
% \caption{Initial configuration of (a) snakuscules (b)hypersnakuscules. Every two neighboring snakuscules or hypersnakuscules are located at a distance $\sqrt{1.5}R$ apart.}
% \label{fig:swarm initials}
% \end{figure*}

\subsection{Parallelizing the process}\label{parallelizing}
Regarding cell localization and counting, 3D snakuscules can be initially placed on the 3D image in a lattice (Figure \ref{fig:all_in_one}b) similar to the 2D case (Figure \ref{fig:snake2D-swarm}b). They evolve independently to segment a nearby spherical structure (blob). However, the higher dimensional integration results in excessive computing time, making a serial implementation impractical for large high resolution images.

% \begin{figure}[!t]
% \centering
% \includegraphics[width=2.5in]{figures/init-final-fit.pdf}
% \caption{A DAPI stained brain tissue slide (a) Initial configuration of snakuscules on the 2D-image, and (b) Final configuration of them.}
% \label{fig:2D conf}
% \end{figure}

%\begin{figure}[!t]
%\centering
%\includegraphics[width=0.9\linewidth]{figures/2D_results.pdf}
%\caption{ (a) A DAPI stained brain tissue slice. (b) The initial configuration and (c) final configuration of %snakuscules on a zoomed region.}
%\label{fig:2D conf}
%\end{figure}

Since the evolution of each contour is completely independent from the others, this process is highly data-parallel and an ideal application for graphic processing units (GPUs). GPUs consist of a large number of parallel processors that can be used for general purpose parallel computing to improve the performance of algorithms that are highly data parallel and can be split into a large number of independent threads. 
A GPU has a local single-instruction on multiple data (SIMD) architecture, making execution of the same program on multiple values extremely efficient. The set of instructions applied on each element is called a kernel \cite{smistad_medical_2015}.
We define our evolutionary instructions as a GPU kernel that can be executed for thousands of snakes in parallel. 

\begin{figure}
\centering
\includegraphics[width=0.9\linewidth]{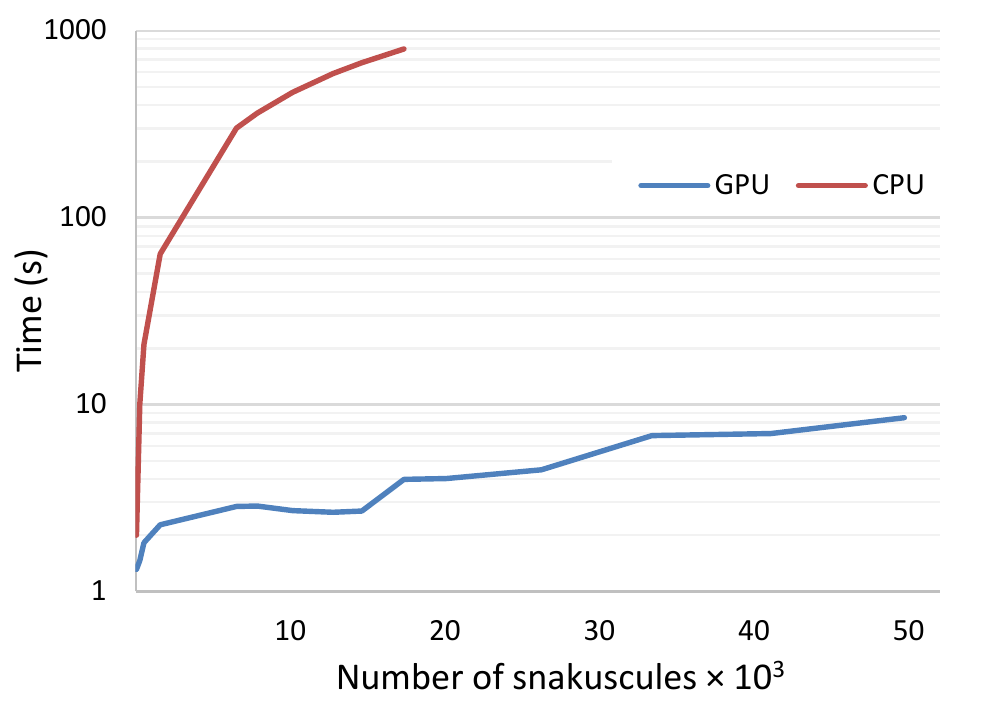}
\caption{Execution time for the method implemented on CPU and GPU. Time axis is plotted on a logarithmic scale.}
\label{fig:profiling-2d}
\end{figure}

For instance, snakuscules are run on various pieces of a 2D DAPI stained rat brain tissue image. The image is a whole rat brain slice with resolution of $350 nm/pixel$. The initial and final snakuscules configurations for a $1000\times1000$ image are shown (Figure \ref{fig:2D conf}). Figure \ref{fig:profiling-2d} illustrates  performance of GPU implementation in comparison to an optimized CPU version. By increasing number of contours, image size, CPU execution time increases significantly, quickly becoming impractical.

The 3D snakuscule is computationally more expensive because of integration over a 3D space using a uniform grid. Therefore, parallel computing using a GPU is employed to assign one contour evolution to one GPU thread. 

\subsection{Monte-Carlo Integration}
In order to further accelerate contour evolution, Monte-Carlo (MC) integration is used. It estimates the integral values using a uniform distribution of randomized samples. In the 2D case, samples are chosen from a uniform distribution inside of a circle with radius $(R + \Delta{R}/2)$. If $r$ and $\theta$ are random numbers in $[0,1]$ and $[0,2\pi)$ respectively; a uniform set of points within the circle with radius $r$ are computed:
\begin{align*}
x&=\sqrt{r} \cos{\theta}\\
y&=\sqrt{r} \sin{\theta}
\end{align*}
MC integration is selected because it provides two advantages over uniform sampling:
\begin{itemize}
    \item Convergence is significantly faster for higher-dimensional data sets, providing an error of $\frac{1}{N}$, regardless of the number of dimensions.
    \item The use of MC sampling allows us to specify a constant number of samples per snake, minimizing branch divergence in the GPU-based SIMD algorithm.
\end{itemize}
One constraint of MC integration is that we are relying on an underlying assumption that the integral is well-behaved (smooth). Given that we expect cell nuclei to be relatively consistent in size, this assumption is well founded. However, it can be mathematically enforced using a low-pass filter that forces the image to be smooth.

\begin{figure}
\centering
\includegraphics[width=0.9\linewidth]{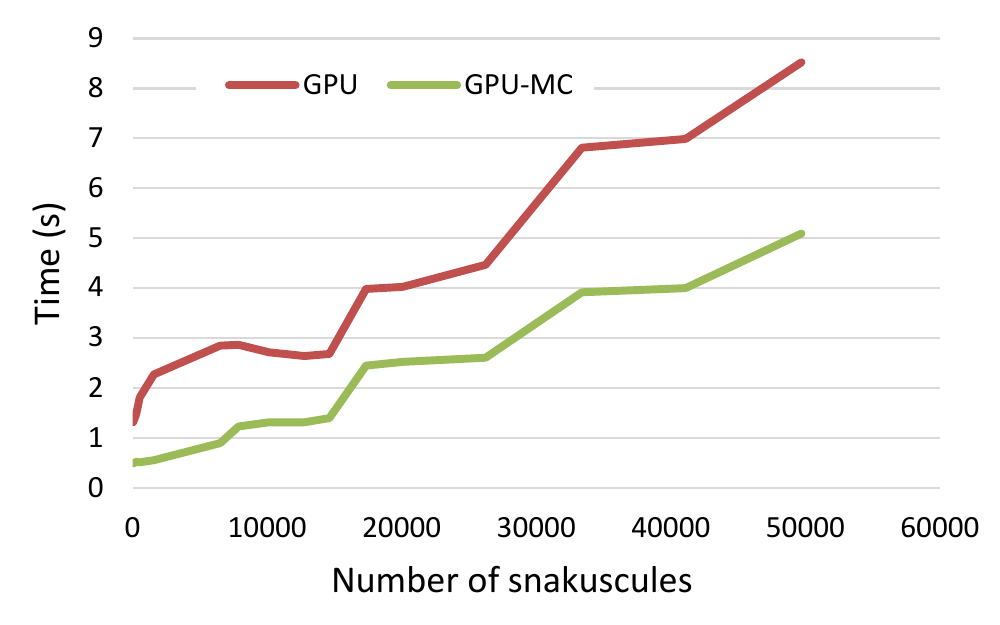}
\caption{Execution time of snakuscules (2D) on GPU with and without Monte-Carlo sampling.}
\label{fig:2D-pro-mc}
\end{figure}

\begin{figure}
\centering
\includegraphics[width=0.9\linewidth]{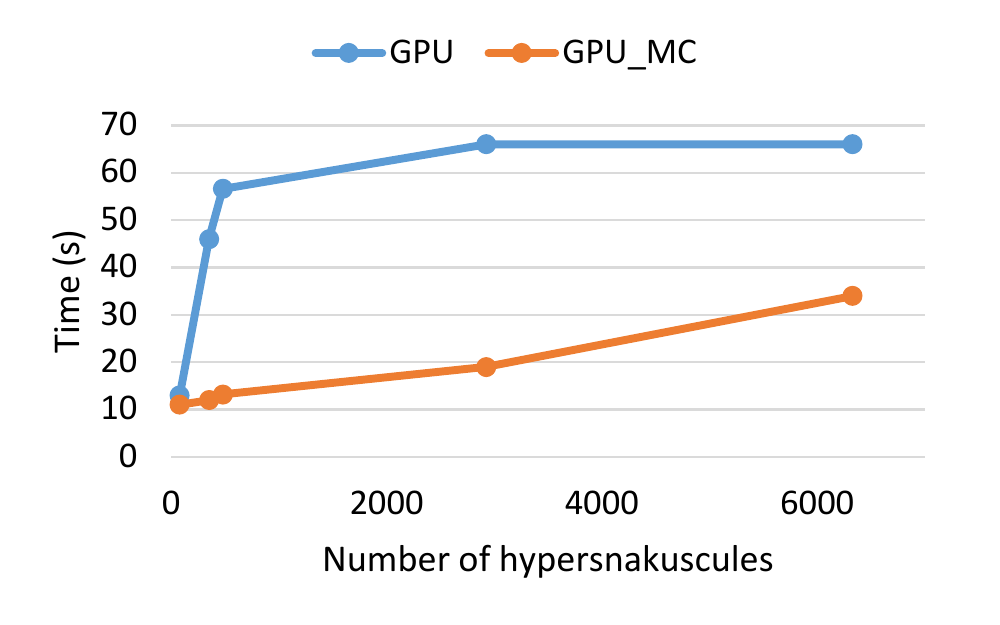}
\caption{Execution time of 3D snakuscules on GPU with and without Monte Carlo sampling}
\label{fig:3D-pro-mc}
\end{figure}

For 3D images, uniform sampling is done within a sphere with radius $(R+\Delta{R}/2)$. Execution time using Monte-Carlo sampling in comparison with the original integration for different number of snakes on the 2D (Figure \ref{fig:2D-pro-mc}) and the 3D (Figure \ref{fig:3D-pro-mc}) images shows significant improvement. As expected, a significantly greater acceleration can be seen in the 3D algorithm, with an $\approx 4X$ gain in performance on average.

\subsection{Parallel Contour Evaluation} \label{further parallel}
In order to improve the GPU efficiency by utilizing more GPU resources, we further parallelize each 3D contour. We instead assign each block to one contour so that threads in that block are responsible for smaller parts of Equation \ref{sum_dEdp} and \ref{sum_dEdq}. For each snake, if MC integration selects $N$ random samples and the CUDA kernel is launched with $T$ threads (the maximum number of threads per block), each thread calculates a portion of the energy (Equation \ref{dEdp.x}-\ref{dEdp.z}) corresponding to $N/T$ spatial locations within the contour. The results are stored in shared memory and combined (Equation \ref{sum_dEdp}-\ref{sum_dEdq}) to calculate the final contour at each iteration. This allows employing more GPU threads to cooperatively walk through a snake evolution process. 

%........................................................................
\section*{Results and Discussion}

% Regarding cell localization and counting, hypersnakuscules can initially congregate on the image closely(Figure \ref{fig:3D_init}). They are evolving independently to lock on their close light blob. It can be done more efficient and faster using GPU-framework in parallel.
In order to find all sphere-like objects in an 3D-image without user interaction, the image is covered by close initial 3D contours. The 3D contours update their current configurations by individually optimizing their energies. Contour evolution is stopped when either (a) they meet maximum number of iterations or (b) they converge.  
Overlapping snakes, as defined by $||\mathbf{c}^\prime-\mathbf{c}^{\prime\prime}||<\max(\mathbf{R}^\prime,\mathbf{R}^{\prime\prime})/\sqrt[3]{2}$, undergo a competition with the lower energy snake surviving. 3D snakuscules with energy greater than a threshold ($\mathbf{E}_0$) are also removed.
\begin{figure}[!t] 
\centering
\includegraphics[width=\linewidth]{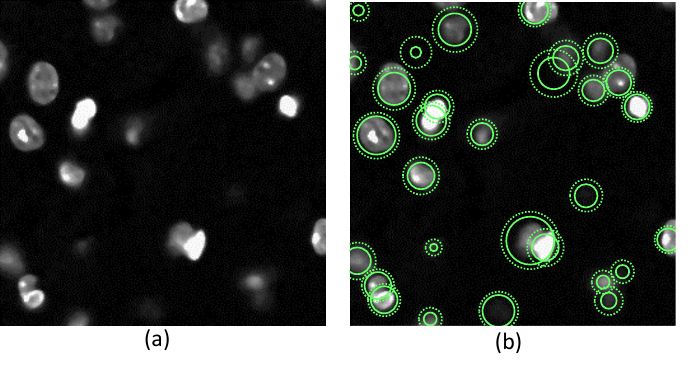}
\caption{(a) A section of DAPI stained mouse hippocampus 3D image and (b) the final configuration of 3D snakuscules on the section.}
\label{fig:3D-init-final}
\end{figure}

Images of the hilus region of the dentate gyrus in the mouse hippocampus were collected with a $40X$ oil objective on a Leica TCS SP8 confocal microscope ($1024\times1024$ pixels; $387.5\times387.5\mu m$). A $405 nm$ laser excited the DAPI signal that was detected between $415-500 nm$. A $1 \mu m$ step size was set for z stack collection of the entire tissue thickness. Acquisition speed was set to 600 Hz, with a 0.75 zoom factor. Raw images for all data analysis were exported as TIFFs. Transgenic mice that model Dravet syndrome with spontaneous seizure onset at postnatal day 15 were housed in a 12 hour light/dark cycle. These mice have a knock-in mutant Scn1A gene containing a nonsense substitution (CgG to TgA) in exon $21$ \cite{ogiwara2007nav1}.
All animal experiments were approved by the Institutional Animal Care and Use Committee of the University of Houston.

We applied our method to the image of size $256\times256\times40$ (Figure \ref{fig:3D-init-final}a). In order to deal with pixel anisotropy, the images were re-sampled to obtain a uniform pixel size. The contours are initialized as a lattice of 3D snakuscules. The 3D snakuscules are evolved and culled using the proposed methods. Figure \ref{fig:3D-init-final}b depicts the final configuration of them on a 2D slice.

%\subsection{Validation and Comparison}
To quantitatively evaluate the performance of our method, 3D snakuscules are considered as either a cell (foreground) or non-cell (background) using a K-nearest neighborhood (KNN) search. The three evaluation parameters, precision ($P_r$), recall ($R_e$) and the F-measure ($F$), are calculated as follows:
\begin{align*}
P_r &= \frac{TP}{TP + FP}\\
R_e &= \frac{TP}{TP + FN}\\
F&= \frac{2P_r R_e}{P_r + R_e} 
\end{align*}

Where the true positive ($TP$) value is number of accurately detected cells, the false positive ($FP$) value is the number of falsely detected cells, and the false negative ($FN$) value is number of undetected cells.

% In this experiment,  we set the initial radius to 15 pixels ($\approx 6\mu m$), the energy threshold $E_0=-3 (E_0\leq0)$ , and the maximum number of iterations to 400. We calculate the precision, recall and F measure as $97\% $, $84\%$ and $90\%$ respectively.

Table \ref{tab:comparison-table} illustrates the performance of MINIS, FARSIGHT, 3D object counter, CellSegm and the proposed method on a DAPI-labeld image with $53$ annotated cells. In this experiment, we set the 3D snake parameters so that initial radius is 15 pixels ($\approx 6\mu m$), the energy threshold $E_0=-3 (E_0\leq0)$ , and the maximum number of iterations is 400. Also, We adjusted the hyperparameters of other methods to optimize their performance on our dataset. The results clearly demonstrate that 3D snakuscules are suitably capable of capturing round cell nuclei, and provide considerable performance advantages over other conventional methods with F-measure of $90\%$ in comparison with that of $82\%$, $74\%$, $62\%$ and $82\%$ for MINIS, FARSIGHT, 3D object counter and CellSegm respectively. Additional advantages include the minimal number of parameters required for initialization.

We also evaluated our algorithm on two publicly available data sets (Figure \ref{diff-datasets}) available at www.celltrackingchallenge.net:
\begin{itemize}
    \item Fluo-N3DH\_CE: Caenorhabditis elegans embryos stained with green flourescent protein (GFP) transfection collected with Plan-Apochromat $63X/1.4$ (oil) objective lense on Zeiss LSM $510$ Meta and the voxel size $0.09 \times 0.09 \times 1.0 \mu m^3$ \cite{mavska2014benchmark , ulman2017objective}.
    
    \item Fluo-N3DH-SIM+: A simulated video from fluorescently labeled nuclei of the HL60 cells stained with Hoescht. It is imaged using Plan-Apochromat $40X/1.3$ (oil) objective with resolution $0.125 \times 0.125 \times 0.2 \mu m^3$  \cite{mavska2014benchmark , ulman2017objective}.
\end{itemize}

\begin{table}[]
\centering
\caption{Performance of different algorithms against manually segmented ground truth through evaluation parameters, precision, recall and Fmeasure}
\label{tab:comparison-table}
\begin{tabular}{llll}
\hline
Method                                                                & Precision     & Recall        & Fmeasure      \\ \hline
MINS                                                                  & 0.75           & 0.90          & 0.82          \\
Farsight                                                              & 0.64          & 0.88          & 0.74           \\
object counter- imagej                                                & 0.59          & 0.66          & 0.62           \\
CellSegm                                                              & 0.66           & 0.90          & 0.82          \\
\textbf{3D snakuscules}                                         & \textbf{0.97} & \textbf{0.84} & \textbf{0.90} \\ \hline
\end{tabular}
\end{table}

\begin{figure}[!t]
\centering
\includegraphics[width=\linewidth]{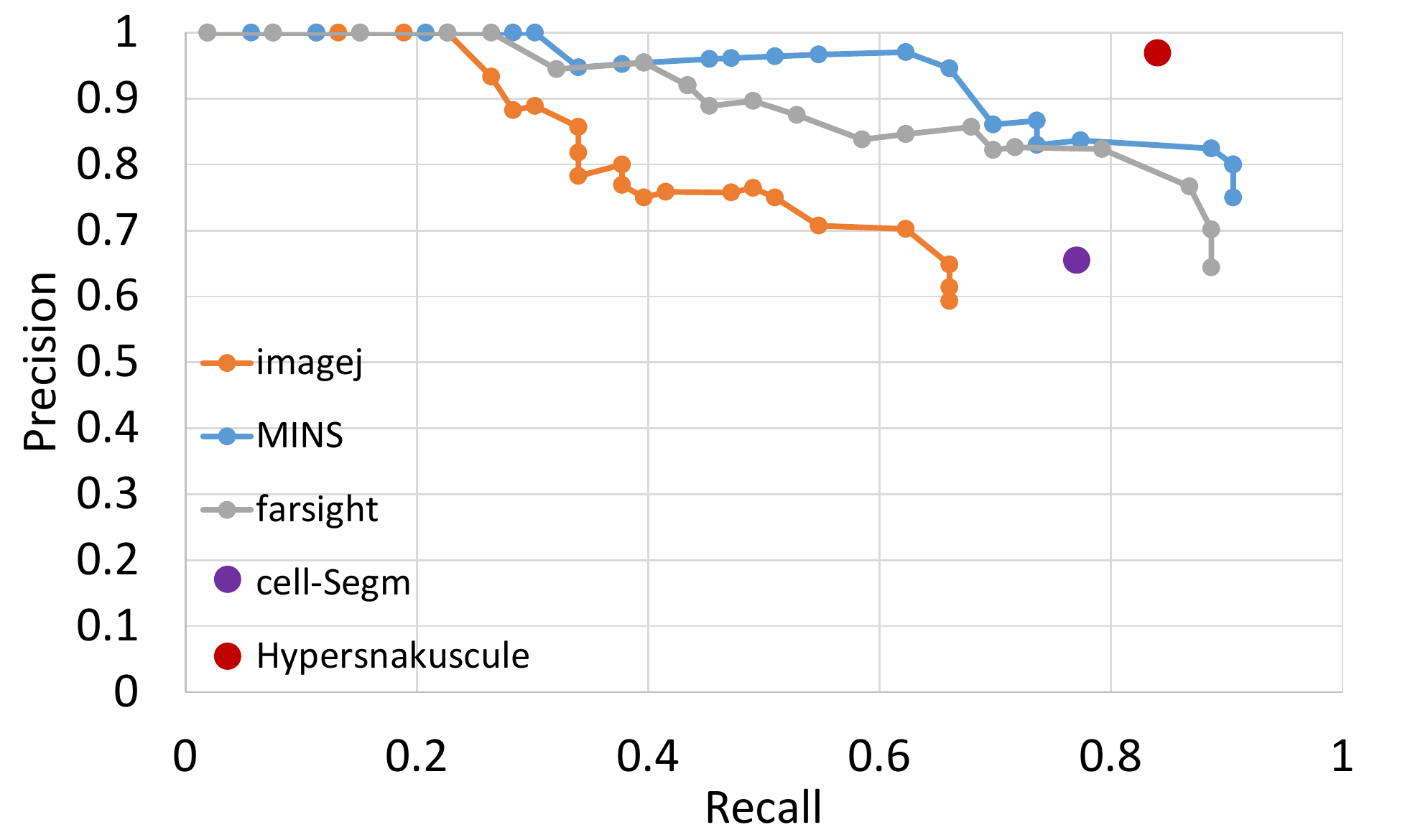}
\caption{Evaluation of different segmentation algorithms on the same DAPI stained image. Our proposed method (3D snakuscule) provides results which matches the ground truth better than others.}
\label{fig:precision-recall}
\end{figure}

\begin{figure}
    \centering
    \includegraphics[width=\linewidth]{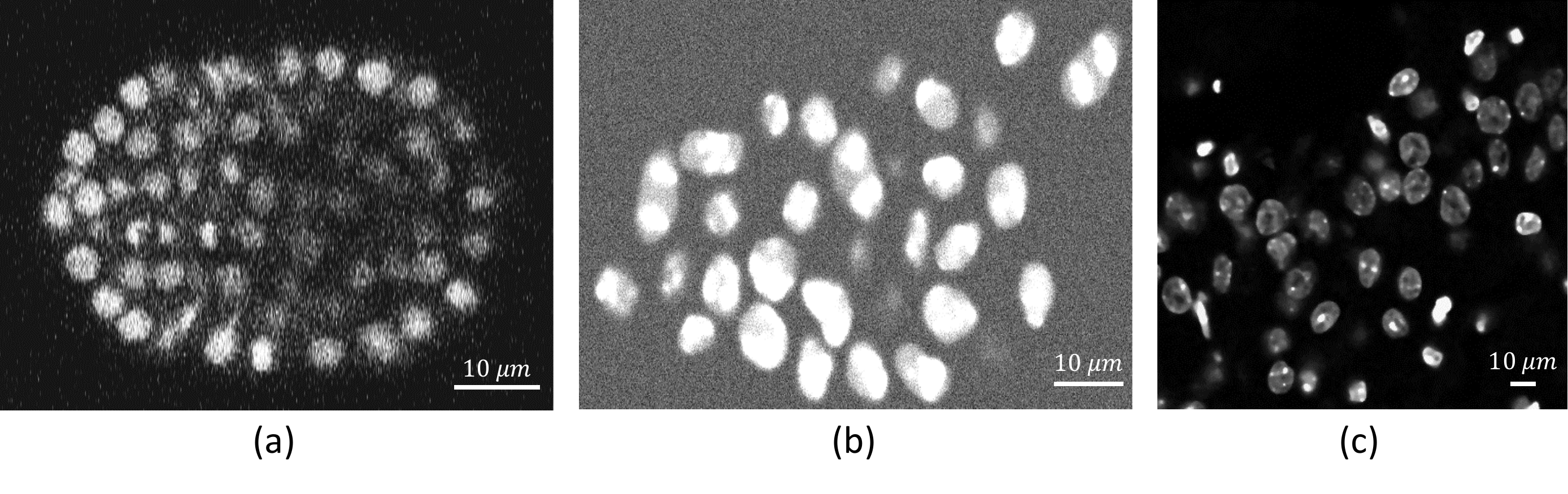}
    \caption{Representing one section of different 3D datasets used for cell detection. (a)Fluo-N3DH\_CE (b)Fluo-N3DH-SIM+ (c) Mouse-Brain}
    \label{fig:diff-datasets}
\end{figure}

\begin{table}[]
\centering
\caption{Performance of 3D snakescule on different datasets with different number of existing cells}
\label{diff-datasets}
\begin{tabular}{lllll}
\hline
Dataset            & \# Cells & Precision & Recall & Fmeasure \\ \hline
Fluo-N3DH\_CE      & 209      & 0.93      & 0.98   & 0.95     \\
Fluo-N3DH-SIM+      & 39       & 0.97      & 0.92   & 0.94     \\
Mouse-Brain(sec.1) & 172      & 0.94      & 0.82   & 0.87     \\
Mouse-Brain(sec.2) & 53       & 0.97      & 0.84   & 0.90     \\ \hline
\end{tabular}
\end{table}

% \subsection{Data Acquisition}
% Images of the hilus region of the dentate gyrus in the mouse hippocampus were collected with a $40X$ oil objective on a Leica TCS SP8 confocal microscope ($1024\times1024$ pixels; $387.5\times387.5\mu m$). A $405 nm$ laser excited the DAPI signal that was detected between $415-500 nm$. A $1 \mu m$ step size was set for z stack collection of the entire tissue thickness. Acquisition speed was set to 600 Hz, with a 0.75 zoom factor. Raw images for all data analysis were exported as TIFFs.
% Transgenic mice that model Dravet syndrome with spontaneous seizure onset at postnatal day 15 were housed in a 12 hour light/dark cycle. These mice have a knock-in mutant Scn1A gene containing a nonsense substitution (CgG to TgA) in exon $21$ \cite{ogiwara2007nav1}.
% All animal experiments were approved by the Institutional Animal Care and Use Committee of the University of Houston.
%------------------------------------------------
\section{GPU Occupancy}
Occupancy is a measure of how many warps the kernel has active on the GPU, relative to the maximum number of warps supported by the GPU.
The graphic processor used in our experiments is GetForce GTX-1070. Theoretical occupancy provides an upper bound while achieved occupancy indicates the kernel's actual performance. When the GPU does not have enough work, resources are wasted.  

\begin{figure*}[t]
\centering
\includegraphics[width=0.9\linewidth]{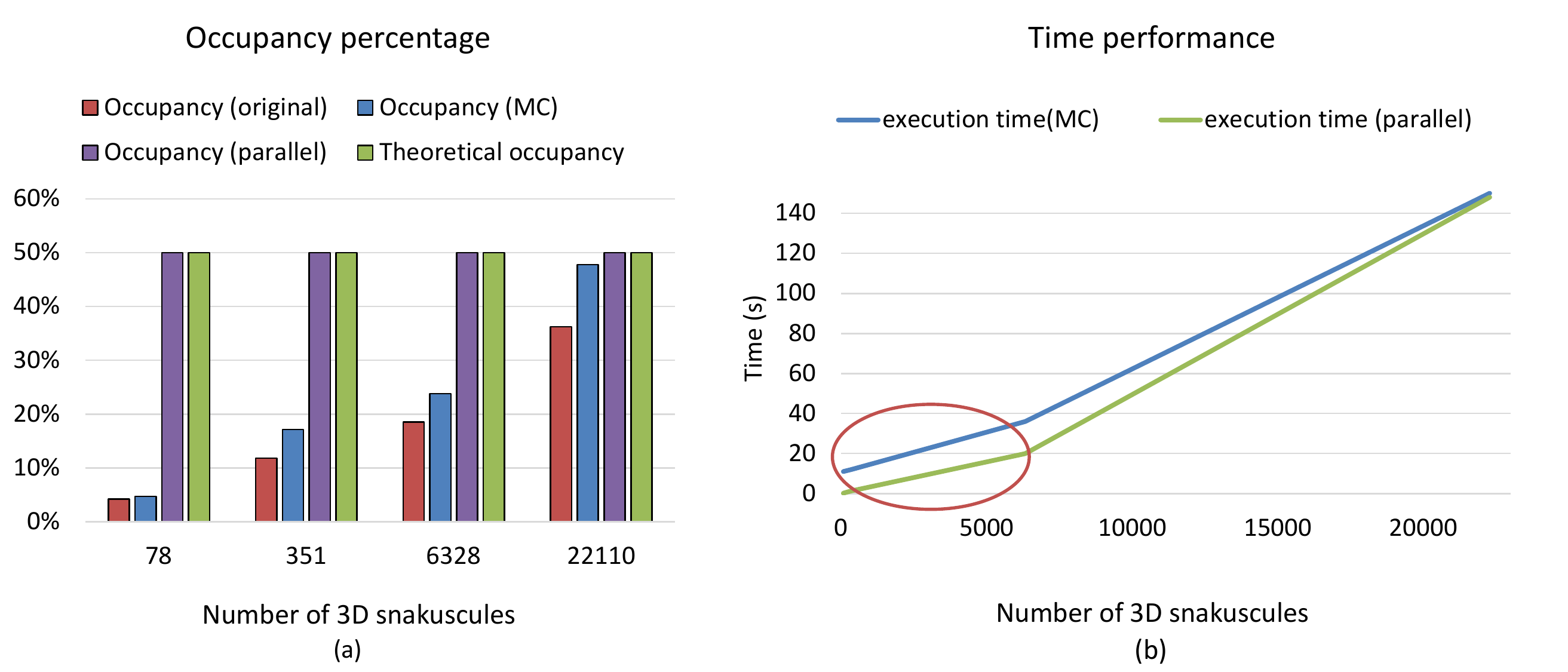}
\caption{(a) Theoretical occupancy, green, and achieved occupancy using different size images (different number of initial 3D snakuscules) with and without Monte Carlo integration, blue and red respectively. Parallel 3D snakuscules provide an achieved occupancy (purple) comparable to the theoretical limit. (b) The diagram shows execution time of the method implemented on GPU using MC integration before and after further parallelization for different number of initiated 3D snakuscules. }
\label{fig:occupancy-time}
\end{figure*}

The theoretical occupancy for our algorithm is $50\%$, limited by the number of registers required for contour evolution. This is a relatively standard theoretical occupancy for complex calculations, however a more rigorous optimization may yield better results in the future. Since any 3D contour is assigned to one GPU thread, the number of utilized threads is equal to the number of initial 3D snakes. Therefore, a small image with a small number of cells occupies fewer resources, resulting in low compute performance that is unable to hide operation and memory latency (Figure \ref{fig:occupancy-time}a). Comparing achieved occupancy with and without Monte-Carlo sampling shows that MC integration improves GPU performance and occupancy by reducing thread stalls. Since there are cells with various sizes in the image, uniform integration takes longer for contours corresponding to bigger cells, resulting in additional stalls. These are mitigated using MC integration. 

Since the reduction in efficiency is due primarily to low occupancy, further parallelizing each 3D contour (Section \ref{further parallel}) allows for the generation of more threads. This provides an achieved occupancy much closer to the theoretical limit (Figure \ref{fig:occupancy-time}a), further reducing processing time (Figure \ref{fig:occupancy-time}b).

% \subsection{Latency Bounds}
% Analyzing the kernel of evolving the parallized 3D snakuscules indicates that the performance limited by the latency of arithmetic or and memory operations (Figure \ref{fig:profiling}b). There are several reasons for latencies (Figure \ref{fig:profiling}c). For evolving a hypersnakuscule in parallel, the preliminary results are stored in shared memory and later combined to update the contour, requiring thread synchronization. Despite performance gains, this is a bottleneck for smaller images. When an input required by the instruction is not available at the time (execution dependency) results in thread stalls. It can be reduced by increasing instruction-level parallelism. Although its portion is about $18\%$ for this kernel, it is much less than that of kernel without further parallelization ($35\%$).

%.....................................................................................
\section{Conclusion}
We developed a 3D blob-detector, based on the miniscule snakes (snakuscule) algorithm, that provides a method for 3D nuclei detection with minimal user interaction. This paper describes a unified formulation of snakuscules in three dimensional space, so that the new cost function is minimized with respect to two points which define the contour. Although the method is initially computationally expensive, it is extremely data parallel and can be efficiently implemented using GPU hardware. A GPU implementation, combined with Monte-Carlo sampling, results in a simple and fast blob detector for large images with numerous cells. Our method requires the specification of a minimum contour size, which is usually readily available for microscopy images. We have illustrated that the GPU implementation and Monte Carlo sampling significantly increase performance, making 3D snakuscules viable for image segmentation. The experimental results demonstrate that the proposed method outperforms state of art methods in overall accuracy.

One major limitation of this algorithm is that a significant portion of the evaluation is devoted to the evolution of snakes that will ultimately be culled. This suggests that any method that reliably places initial contours could significantly increase snakuscule performance. A more optimal placement of initial contours could significantly improve performance beyond what we were able to achieve with a lattice. However, methods such as iterative voting and Laplacian of Gaussian blob detection resulted in reduced accuracy when tested. Therefore more advanced algorithms, such as dynamic culling or insertion of contours during evolution, may be a better approach.

In addition, further optimization of the evolution kernel to increase theoretical occupancy limited by register usage could double performance.

%........................................................................

% or


% \appendices

% % you can choose not to have a title for an appendix
% % if you want by leaving the argument blank
% \section{}

\section*{Acknowledgment}

This work was funded in part by the Cancer Prevention and Research Institute of Texas (CPRIT) \#RR140013, the National Institutes of Health / National Library of Medicine (NLM) \#4 R00 LM011390-02, as well as the National Science Foundation I/UCRC BRAIN Center \#1650566. The authors would like to thank Dr. Leigh Leasure and Dr. Amy Sater for the use of their staining reagents and access to microscopy equipment.

\ifCLASSOPTIONcaptionsoff
  \newpage
\fi



\bibliographystyle{IEEEtran}
\bibliography{MyLibrary.bib}
\end{document}